\documentclass[aps, prl, preprint, superscriptaddress, nobibnotes, amsmath, amssymb]{revtex4-2}

\usepackage{graphicx}
\usepackage{color}

\usepackage{hyperref}
\hypersetup{
    colorlinks=true, 
    linktoc=all,     
    linkcolor=blue,
    citecolor=blue,
    urlcolor=black,
    breaklinks=true
    }  
\usepackage[all]{hypcap}

\begin{document}
\title{Multiple Magnetoelectric Plateaux in Polar Magnet Fe$_2$Mo$_3$O$_8$}

\author{Qian~Chen}
\author{Atsushi Miyake}
\affiliation{Institute for Solid State Physics, The University of Tokyo, Kashiwa, Chiba 277-8581, Japan}

\author{Takashi Kurumaji}
\affiliation{Department of Advanced Materials Science, The University of Tokyo, Kashiwa 277-8561, Japan}

\author{Keisuke Matsuura}
\affiliation{RIKEN Center for Emergent Matter Science, Wako 351-0198, Japan}

\author{Fumitaka Kagawa}
\affiliation{RIKEN Center for Emergent Matter Science, Wako 351-0198, Japan}
\affiliation{Department of Physics, Tokyo Institute of Technology, Tokyo 152-8551, Japan}

\author{Shin Miyahara}
\affiliation{Department of Applied Physics, Fukuoka University, Jonan-ku, Fukuoka 814-0180, Japan}

\author{Yoshinori Tokura}
\affiliation{RIKEN Center for Emergent Matter Science, Wako 351-0198, Japan}
\affiliation{Tokyo College and Department of Applied Physics, The University of Tokyo, Hongo, Tokyo 113-8656, Japan}

\author{Masashi Tokunaga}
\affiliation{Institute for Solid State Physics, The University of Tokyo, Kashiwa, Chiba 277-8581, Japan}

\maketitle

{\bf Abstract}

The magnetization and electric polarization of a polar antiferromagnet Fe$_2$Mo$_3$O$_8$ are studied up to 66 T for spin-saturation magnetic fields applied along the polar axis. 
The magnetization process at 1.4 K exhibited multistep structures below the saturation field of 65 T. 
The electric polarization along the polar axis exhibits a similar multistep behavior with a total change of 1.2 $\rm{\mu}C/cm^{2}$.  
A combined triangular-lattice antiferromagnetic model with strong Ising-type spin anisotropy reproduces this multistep magnetoelectric (ME) effect. 
The exchange striction mechanism explains the remarkable ME response in the two sub-lattice type-I multiferroic materials.
These results and interpretation demonstrate a method for realizing multistage magnetoelectric effects in hybrid spin systems.

\newpage
In recent years, the rapid development of information technologies has driven the search for new materials and mechanisms for electronic devices.
In particular, ``multiferroic" materials, which were first discovered over half a century ago \cite{dzyaloshinskii1960magneto,astrov1960magnetoelectric}, have become a promising candidate for future low-power memory devices. 
The industrial demands have resulted in increasing scientific and industrial interst in ``multiferroics" since 2000 \cite{schmid1994introduction,hill2000there,eerenstein2006multiferroic,tokura2007multiferroics,tokura2010multiferroics,vopson2015fundamentals}. 
The prominent magnetoelectric (ME) effect in TbMnO$_3$ revealed type-II multiferroic materials in which the ferroelectricity originates from their magnetic order \cite{kimura2003magnetic}. 
However, recent studies revealed that magnetic transitions in type-I multiferroic materials, in which the ferroelectricity or polarity is of different origin from the magnetism, can also have a significant ME effect \cite{dho2006large}.

The honeycomb magnet Fe$_2$Mo$_3$O$_8$ is one of the type-I multiferroic materials, which belongs to the family of ternary transition-metal molybdenum oxides, possessing a polar hexagonal structure (pyroelectric space-group \textit{P}-6$_3$\textit{mc}) \cite{mccarroll1957some} with \textit{c}-axis crystallographic polarity.
In Fe$_2$Mo$_3$O$_8$, the Fe$^2$$^+$ ions are located in tetrahedrally coordinated sites (A-sites) as well as in octahedrally coordinated sites (B-sites) of oxygen and are interleaved in the $ab$ plane to form honeycomb layers (Fig. 1(a)). 
The Mo$^4$$^+$ ions form non-magnetic spin-singlet trimers \cite{cotton1964metal} and separate each Fe$^2$$^+$ honeycomb layer along the $c$ axis. 
The A and B sites on adjacent layers occupy opposite positions. 
Below a N\'{e}el temperature of 60 K, the spins of the Fe$^2$$^+$ ions ($S=2$) are aligned in a collinear antiferromagnetic (AFM) order \cite{mcalister1983magnetic}. 
In this AFM state, the spins in the nearest inter-plane A-B sites point in the same direction, whereas those in the nearest intra-plane A-B sites take opposite orientations (Fig. 1(b)). 
A previous study \cite{ghara2022magnetization, varret1972etude} reported Ising-like anisotropy in B-site iron spins. 
Because the magnetic moment at the B site (\textit{m}$_{\rm{B}}$) is slightly larger than that at the A site (\textit{m}$_{\rm{A}}$) \cite{bertrand1975structural}, the application of magnetic fields along the polar axis stabilizes the ferrimagnetic (FiM) state.
In this FiM state, all the spins of the A and B sites are aligned antiparallel and parallel to the magnetic field, respectively.
The competition between the AFM and FiM orders can be finely tuned by chemical doping \cite{bertrand1975structural}. 
Fe$_2$Mo$_3$O$_8$ and its Zn-doped compounds exhibit a linear ME \cite{kurumaji2015doping}, optical ME \cite{kurumaji2017optical}, electromagnon excitations \cite{kurumaji2017electromagnon}, and significant thermal Hall effects \cite{ideue2017giant}. 
In addition, the giant ME response \cite{kurumaji2015doping,wang2015unveiling} reveals the nature of the exchange-striction mechanism \cite{arima2007ferroelectricity}.

In this study, we investigate the high-field properties of honeycomb polar magnet Fe$_2$Mo$_3$O$_8$. 
Multistep magnetization and the corresponding giant ME responses are observed at low temperatures when a magnetic field is applied along the polar axis. 
We interpret the experimental findings using a combined triangular lattice model with Ising-type anisotropy and the exchange-striction mechanism.

Single crystals of Fe$_2$Mo$_3$O$_8$ (samples K1, K2, and K3) were grown by the chemical vapor transport method \cite{strobel1983growth}. 
The crystal orientations were confirmed using X-ray Laue diffraction measurements.
Pulsed-magnetic fields of up to 66 T with durations of 4 to 36 ms  were generated by several types of non-destructive pulse magnets provided by the International MegaGauss Science Laboratory at the Institute for Solid State Physics, the University of Tokyo. 
The magnetization ($M$) in pulsed-fields  was measured using the conventional induction method with coaxial pick-up coils.
The absolute value of the magnetization was calibrated by magnetization measurements using a commercial superconducting quantum interference device magnetometer system (MPMS; Quantum Design).
The relative change in electric polarization ($\Delta$$\it{P}$) with respect to the initial values at zero field was measured by integrating the polarization currents induced by the magnetic fields \cite{mitamura2007dielectric}. 
The electrodes were constructed using gold wires with carbon-based paste painted on two flat surfaces perpendicular to the $c$ axis. 
The sign of the polarization taken from \cite{kurumaji2015doping}, is defined as the spin-driven contribution appearing negative below ~60 K. 
The extrinsic background caused by the large time-derivative of the magnetic fields was evaluated by reversing the polarity of the magnetic fields above the N\'{e}el temperature and removed from each data.

\begin{figure}[!ht]
\centering
\includegraphics[scale=1]{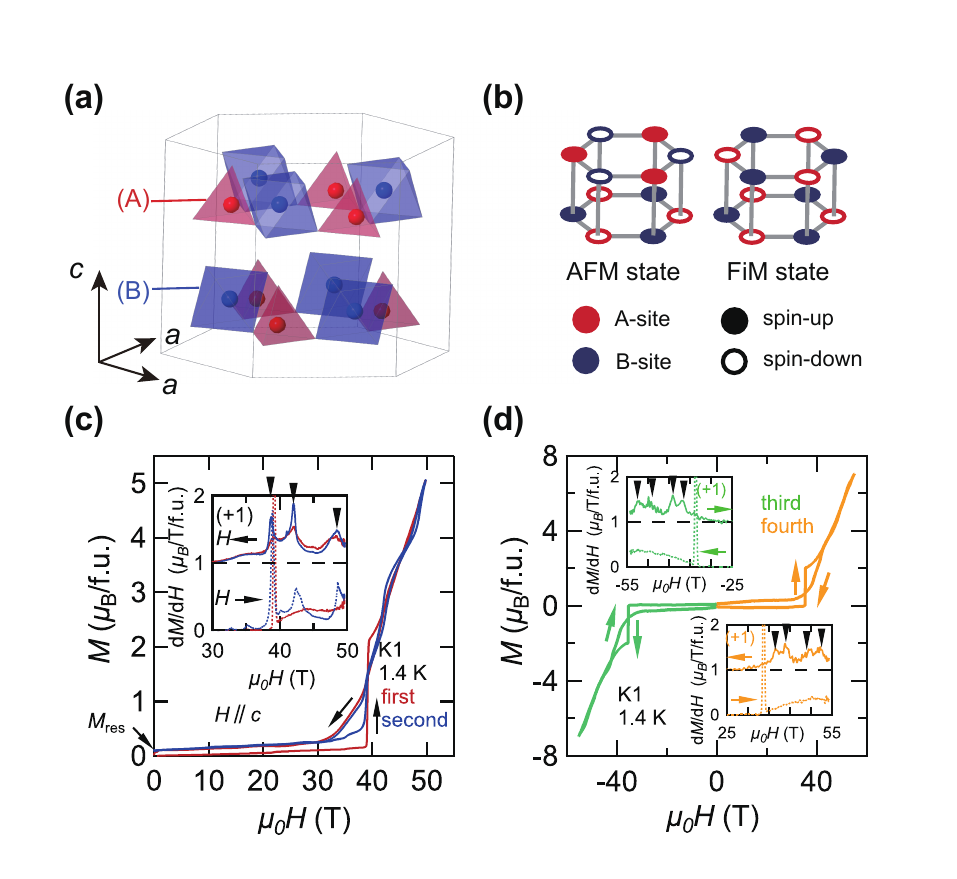} 
\caption{
    (a) Schematic arrangement of FeO$_4$ and FeO$_6$ in a magnetic unit cell of Fe$_2$Mo$_3$O$_8$. 
    The red and blue circles represent Fe ions in the A- and B-sites, respectively.
    (b) Schematic of the AFM and FiM states. 
    (c) Virgin (red) and secondary (blue) magnetization curves of Fe$_2$Mo$_3$O$_8$ at 1.4 K in pulsed magnetic field (durations of 10 ms) applied along the polar axis.  
    (d) Magnetization curves after reversing the polarity of magnetic fields (durations of 36 ms). 
    The insets in (c) and (d) show the field derivative of the magnetization curves with corresponding colors. 
}   
\label{fig:fig1}
\end{figure}

As shown in Fig. \ref*{fig:fig1}(c), we measured the magnetization of Fe$_2$Mo$_3$O$_8$ in pulsed magnetic fields along the $c$ axis at 1.4 K. 
The magnetization increases steeply at approximately 40 T during the field-increasing process of the virgin trace.
Although there was a relatively smooth change in the further field-increasing process of the virgin trace, a clear wiggling behavior emerged in the field-decreasing process and the secondary trace. 
In the field-derivative curve panel (inset), we obtained three peak structures at approximately 39, 41, and 48 T below 50 T as marked by black triangles. 
A finite residual magnetic moment ($M_{\rm{res}}$) exists in the virgin cycle at this temperature when the magnetic field decreases to zero.   
From the secondary measurements, the magnetization curves exhibit wiggling behaviors in both field-increasing and decreasing processes while maintaining this temperature.
On application of a negative field afterward, the entire process was reset. 
We observed wiggling traces after a smooth change in the field-increasing process as shown in Fig. \ref*{fig:fig1}(d). 
When the magnetic field direction is reversed, the magnetization exhibits an antisymmetric character.

\begin{figure}[!ht]
    \centering
    \includegraphics[scale=1.2]{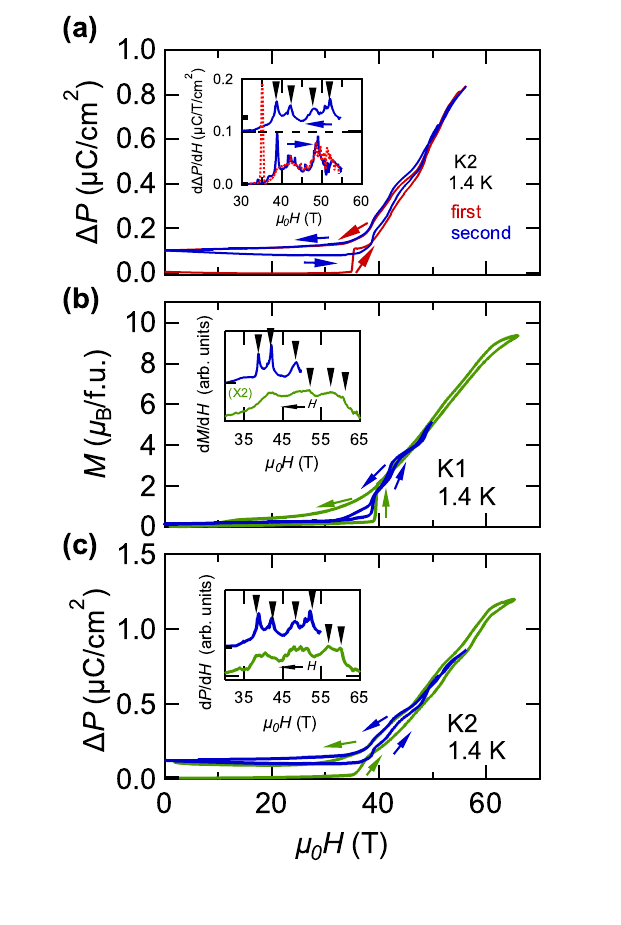} 
    \caption{
      (a) Virgin (red) and secondary (blue) electric polarization curves of Fe$_2$Mo$_3$O$_8$ at 1.4 K in pulsed magnetic field (durations of 36 ms) applied along the polar axis.
      (b) Magnetization in magnetic fields of up to 66 (green) and 50 T (blue). 
      (c) Electric polarization in magnetic fields of up to 66 (green) and 56 T (blue). 
      The insets show the field derivative of the magnetization and electric polarization curves with corresponding colors in the field-decreasing process. 
    }   
    \label{fig:fig2}
    \end{figure}

Next, we measured the field-induced change in electric polarization along the $c$ axis at 1.4 K (Fig. 2(a)). 
History-dependent multistep behavior was also observed in the $P-H$ curves. 
The $P-H$ profiles in the field-increasing process of the virgin trace is relatively smooth, similar to the case of the $M-H$ measurements. 
In the inset of Fig. \ref*{fig:fig2}(a), four peak structures are observed in the field derivative curves below 55 T. 
Here, the zero point of $\Delta$$\it{P}$ is defined by the initial value of the electric polarization in the virgin trace. 

To clarify the behavior up to the saturation field above 60 T, we measured the magnetization and electric polarization using a short-pulse magnet (with a duration of 4 ms). 
As indicated by the green curve in Fig. \ref*{fig:fig2}(b), the magnetization reaches approximately 9.2 $\mu_{B}$/f.u. at 66 T, which is close to the expected saturation moment ($m_{\rm{A}}+m_{\rm{B}}=\rm{4.21\, \mu}_{B}/f.u.+4.85\, {\mu}_{B}/f.u$) of the material \cite{bertrand1975structural}. 
In this fully spin-polarized (FSP) state, all the spins of the A- and B-site iron ions align parallel to the magnetic field.  
Moreover, the total change in the electric polarization amounts to 1.2 $\rm{\mu}C/cm^{2}$ at 66 T as shown in Fig. \ref*{fig:fig2}(c). 
From the derivative curves of the field-decreasing process, we obtained six peak structures between 35 and 65 T (insets).  
The field positions of the peaks between the polarization and magnetization measurements almost coincide. 
Nevertheless, the peaks are more prominent in mid-pulse magnets (Fig. 1(c) and Fig. 2(a), duration of 10 to 36 ms); therefore, it takes sufficient time to stabilize the intermediate states discussed later. 

\begin{figure}[!ht]
    \centering
    \includegraphics[scale=0.8]{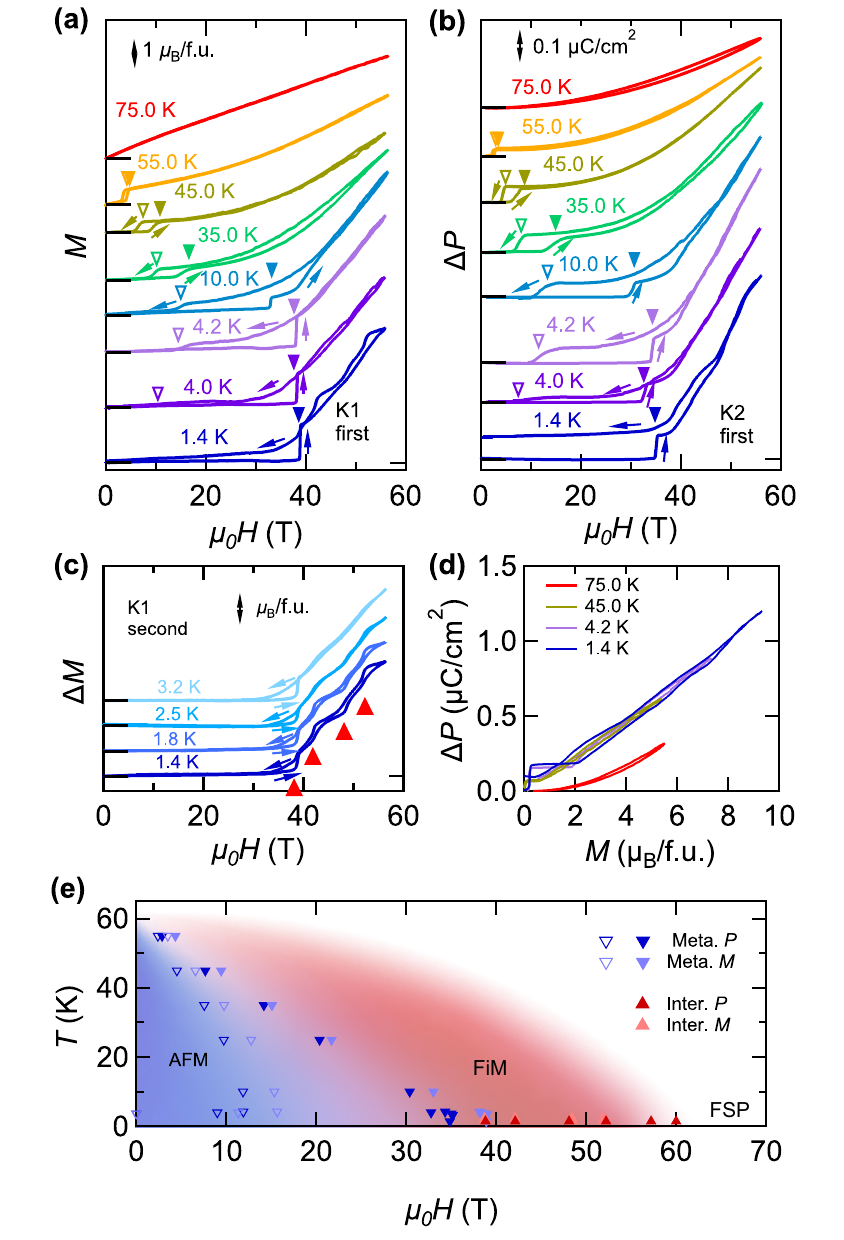} 
    \caption{
        Magnetic field dependence of (a)$M$ and (b)$\Delta P$ at various temperatures. 
        The data were vertically offset for clarity. 
        (c) Temperature dependence of multistep structures in magnetization below 3.2 K. 
        (d) Relationship between $\Delta P$ and $M$ at different temperatures. 
        (e) Magnetic phase diagram in the $T$-$H$ plane. 
        The closed (open) inverted triangles represent critical fields of AFM$\rightarrow$FiM (FiM$\rightarrow$AFM) transitions. 
        The red triangles represent peak fields in the $dM/dH$ and $dP/dH$ curves.  
        The Meta. $P$ ($M$) and Inter. $P$ ($M$) denote the transition fields between AFM and FiM states, and the transition fields between intermediate states in the $P$ ($M$) curves, respectively.
        }   
    \label{fig:fig3}
    \end{figure}

We also investigated the field dependence of the magnetization and polarization at various temperatures (Figs. \ref*{fig:fig3}(a) and 3(b)). 
The field at the first steep increase in $M$ and $\Delta P$ shifted to a lower value with increasing temperature and merged with the reported transition fields from the AFM state to the FiM state at high temperatures \cite{kurumaji2015doping}. 
Therefore, we attribut all the anomalies marked by triangles to transitions between the AFM and FiM states.
Below approximately 4 K, we observed a finite residual magnetic moment and electric polarization as the magnetic field returned to zero. 
In addition, the wiggling behavior gradually blurred with increasing temperature and was barely visible at 3.2 K as shown in Fig. \ref*{fig:fig3}(c). 
In Fig. 3(d), the $\Delta P$ change linearly with the $M$ below the N\'{e}el temperature of 60 K. 
In the paramagnetic region, $\Delta P$ is proportional to the square of $M$; see the red curve at 75 K in Fig. 3(d). 
We plotted the peak positions of the $dM/dH$ and $dP/dH$ curves in the temperature-field plane (Fig. \ref*{fig:fig3}(e)). 
The multistep behavior only appears when the remnant FiM state exists below approximately 3 K. 
Although there is a finite discrepancy in the transition field between the AFM and FiM phases determined by $P$ (sample K2) and $M$ (sample K1) at low temperatures, we confirmed that this discrepancy originates from the slight sample dependence. 
Simultaneous measurements of $M$ and $\Delta P$ in one sample (K3) comfirmed their coincidence at 1.4 K (see supplementary figure).
Moreover, multistep behavior appears at the same fields for all samples. 

In the following, we discuss the origin of the observed multi-step magnetization process. 
As mentioned earlier, we consider a phase transition from the AFM state to the FiM state at the first magnetization step. 
The FSP state is realized under a magnetic field of 66 T. 
Figures 4(a) and 4(b) schematically show the respective spin configurations in these states. 
The red and blue circles denote the A- and B-site ions, respectively, and the open and closed circles represent down and up spins, respectively. 
In both states, the B-site spins align parallel to the magnetic field, whereas the A-site spins are antiparallel and parallel in the FiM and FSP states, respectively. 
Our experimental results suggest the emergence of five intermediate states during the reorientation of the A-site spins.

If we consider only the A-site spins, they form a triangular lattice in the $ab$ plane. 
The magnetization processes of triangular lattice antiferromagnets are usually expressed by three magnetic sub-lattices.
Because the crystallographic unit cell of the present material contains ions in adjacent layers, we consider six magnetic sub-lattices for both the A- and B-site spins. 
Considering its strong Ising-type nature \cite{ghara2022magnetization}, we express the spin Hamiltonian of this material as,
\begin{equation}
    \begin{aligned}
    \mathcal{H} =&-{J_{1}}\sum_{i\in \rm{A}}s_{i}(\sigma_{i+e_{1}}+\sigma_{i+e_{2}}+\sigma_{i+e_{3}})-{J_{\rm{A}\rm{B}}}\sum_{i\in \rm{A}}s_{i}(\sigma_{i-c}+\sigma_{i+c})\\
    &-{J_{\rm{A}\rm{A}}}\sum_{i\in \rm{A}} s_{i}(s_{i+e_{1}+c}+s_{i+e_{2}+c}+s_{i+e_{3}+c})-{J_{\rm{B}\rm{B}}}\sum_{i\in \rm{B}}\sigma_{i}(\sigma_{i+e_{1}+c}+\sigma_{i+e_{2}+c}+\sigma_{i+e_{3}+c})\\
    &-{J_{2\rm{A}}}\sum_{i\in \rm{A}} s_{i}(s_{i+a_{1}}+s_{i+a_{2}}+s_{i+a_{3}})-{J_{2\rm{B}}}\sum_{i\in \rm{B}}\sigma_{i}(\sigma_{i+a_{1}}+\sigma_{i+a_{2}}+\sigma_{i+a_{3}})\\
    &-\mu_{0}H(\sum_{i\in \rm{A}}m_{\rm{A}}s_{i}+\sum_{i\in \rm{B}}m_{\rm{B}}\sigma_{i}),
    \end{aligned}
\end{equation}
where $s_{i},\sigma_{i}=\pm1$ denote the normalized Ising variables of the A- and B-site iron spins, whereas $m_{\rm{A}}$ and $m_{\rm{B}}$ represent the magnitude of the moments, respectively.
Here, $e_{n}, a_{n}, c$ here describe the relative locations of the nearest intra-plane neighbor sites, second nearest intra-plane neighbor sites, and nearest inter-plane neighbor sites, respectively (Fig. 4(c) and 4(d)). 
Here, we consider the exchange interactions between the nearest intra-plane A-B sites ($J_{1}$), nearest inter-plane A-B sites ($J_{\rm{AB}}$), second nearest intra-plane A-A sites ($J_{\rm{2A}}$), second nearest intra-plane B-B sites ($J_{\rm{2B}}$), nearest inter-plane A-A sites ($J_{\rm{AA}}$) and nearest inter-plane B-B sites ($J_{\rm{BB}}$) together with the Zeeman term.

Let us consider the sequential spin-flipping process from the AFM state to the FSP state. 
Here, we label the state as $\rm{N}\it{n}$, where the integer $n = 0 - 6$ represents the number of up-spins in the six magnetic sublattices in the A-sites, that is, N0 and N6 for the FiM and FSP states, respectively.
As an example, in Figs. 4(c) and 4(d), we illustrate intra-plane and inter-plane spin configurations in the N1 state.
Sequential transition can be reproduced by choosing the appropriate parameters for this Hamiltonian. 
In this model, the magnitude of magnetization in the N$n$ state is given by,
\begin{equation}
    M_{\rm{N}\it{n}}=m_{\rm{B}}-(1-\frac{n}{3})m_{\rm{A}}.
\end{equation}
From the $M$-$H$ curves at 1.4 K, we determined the magnetic moments $m_{\rm{A}}=4.3$ $\rm{\mu_{B}/Fe}$ and $m_{\rm{B}}=4.9$ $\rm{\mu_{B}/Fe}$, which are close to the values determined by neutron diffraction measurements \cite{bertrand1975structural}.
Using the reported value of $J_{\rm{AB}}=0.100$ $\rm{meV}$ \cite{ghara2022magnetization}, we determined the other parameters as  $J_{\rm{1}}=-4.17$ $\rm{meV}$, $J_{\rm{AA}}=-0.140$ $\rm{meV}$, $J_{\rm{BB}}=0.007$ $\rm{meV}$ and $J_{\rm{2A}}=-0.296$ $\rm{meV}$. 
Here, we cannot fix the value of $J_{\rm{2B}}$ because it only causes identical energy shifts for all states considered.
The details of the calculations are described in the Supplementary Materials S2. 
As the transition field from the AFM state to the FiM state is hysteretic and sensitive to the exchange interaction parameters, a finite discrepancy is observed in the  transition field between different samples.

Here, we note the multiple possible spin configurations in the N2, N3, and N4 states. 
If we express the spin configuration by (sgn($s_{1}$), sgn($s_{2}$), sgn($s_{3}$))/(sgn($s_{4}$), sgn($s_{5}$), sgn($s_{6}$)), two N2 states $(-,-, +) / (-, -, +)$ and $(+, +, -) / (-, -, -)$ have different energies. 
For the parameters that reproduce the multi-step $M$-$H$ curve, we found the lowest energy states for N2, N3, and N4 states to be $(-, -,+) / (-, -, +)$, $(+, +, -) / (-, -, +)$, and $(+, +, -) / (+, +, -)$, respectively.
Notably, there are other states including four or two A-site sublattices state in one magnetic unit cell. 
We can rule out this four sublattice state by considering the frustration in the exchange striction effect around the B-site ions (see Supplementary Materials S3).

Next, we focus on the ME effect. 
The ME effect at low fields in this material was explained by the $g$-factor mechanism \cite{kurumaji2015doping}. 
To reproduce the ME effects observed in the FiM states using the similar quadratic function of the magnetic field, we have to introduce unnatural temperature dependence of the coefficients. 
The $g$-factor model may explain the ME effect at low fields, as seen below 40 T in Fig. 2(a), but we discuss the other origin for the gigantic ME effects observed in the FiM states below.
Among the three major mechanisms \cite{tokura2014multiferroics}, we can rule out the inverse Dzyaloshinskii-Moriya interaction because of the collinear spin configuration in the present material. 
In addition, the $p$-$d$ hybridization mechanism may not be relevant because a $180^{\circ}$ spin-flip does not cause any change in the electric polarization in this mechanism. 
Therefore, we consider the exchange striction mechanism \cite{sergienko2006ferroelectricity} to describe the ME coupling in Fe$_2$Mo$_3$O$_8$. 

Here, we express the electrical polarization caused by the exchange striction mechanism using the following equation, 
\begin{equation}
    P^{c}=\frac{1}{V}\left[ \sum_{i\in \rm{A}}\Pi_{\parallel } s_{i}(\sigma_{i+e_{1}}+\sigma_{i+e_{2}}+\sigma_{i+e_{3}}) +\sum_{i\in \rm{A}}(\Pi_{\perp }^{+} s_{i}\sigma_{i+c/2}+\Pi_{\perp }^{-} s_{i}\sigma_{i-c/2})\right],
\end{equation}
where $V$ is the volume of the crystallographic unit cell, $\Pi_{\parallel}$ and $\Pi_{\perp }^{\pm}$ denote contributions from the adjacent intraplane and interplane AB spin arrangements, respectively.
We distinguish between $\Pi_{\parallel}$ and $\Pi_{\perp }^{\pm}$ by noting that the tetrahedron is asymmetric at the top and bottom. 
Using this model, the electric polarization changes by $2(\Pi_{\perp }^{+}+\Pi_{\perp }^{-})$ per A-site in the transition from the AFM to the FiM phase and by $6\Pi_{\parallel}$ to the FSP phase.
As shown in Fig. 2(c), the latter is ten times larger in the experiment, indicating that the intra-plane contribution to the change in electric polarization is dominant.
When B-sublattice spins are fixed to the direction of the external magnetic field, $\sigma_{i}=1$, the electric polarization in Eq.3 is proportional to the magnetization at A-sublattice, $P^{c}\varpropto\sum(3\Pi_{\parallel}+\Pi_{\perp }^{+}+\Pi_{\perp }^{-})s_{i}$. 
Such a relationship is consistent with the experimental observation at low temperatures shown in Fig. 2(d). 

The solid blue lines in Figs. 5(a) and 5(b) show the field-dependence of magnetization and electric polarization calculated for multistep spin configurations in Fig. 5(a) using Eqs. 2 and 3.
The red and green-dashed lines in the figure are the corresponding experimental results.
Since the three parameters in Eq. 3 cannot be determined independently from the information obtained in this experiment, we tentatively use $\Pi_{\parallel}=2.7\times10^{-31}$ C$\cdot$m, $\Pi_{\perp }^{+}+\Pi_{\perp }^{-}=-7.2\times10^{-32}$ C$\cdot$m.  
As can be seen in the figure, the experimental and theoretical results are in generally good agreement, suggesting the validity of the model. 
In particular, the locations of the transition fields are reasonably reproduced using the present parameters, as shown in the inset of Fig. 5(b). 

We can reproduce the multistep $M$-$H$ and $P$-$H$ curves using this triangular-lattice Ising model, assuming fully polarized B-site spins from the FiM state to the FSP state. 
However, at the first magnetization process from the AFM state, we have to flip all the B-site spins taking over strong Ising-type anisotropy. 
A sequential spin-flip process was proposed at the boundary region between the AFM and N0 states \cite{matsuura2023thermodynamic,ghara2022magnetization}.
Such gradual spin-flip may cause smooth changes in the $M$-$H$ and $P$-$H$ curves also in the FiM state, as observed in the virgin traces shown in Figs. 1 and 2. 

In type-II multiferroic materials, magnetic order that breaks spatial inversion symmetry is essential to generate electric polarization. 
However, type-I multiferroic materials, which do not have inversion symmetry even in a paramagnetic state, can host the magnetoelectric effect without symmetry-lowering of the magnetic order. 
As in the present material, the electric polarization originating from the exchange striction mechanism can be inverted between parallel and antiparallel spin configurations. 
Since exchange striction mechanisms are known to cause significant magnetoelectric effects, our results reveal the application potential (for magnetic sensor, data storage, etc.) of type-I multiferroic materials.

Furthermore, we demonstrate that when one sublattice spin is fixed in a ferrimagnet, we can realize the multistage magnetoelectric states through the sequential flipping process of the other sublattice spin. 
Although the present material exhibits the multistage behavior only at low temperatures and high magnetic fields, the basic idea sheds new light on the possibility of future applications of multi-state ME devices.

Recent several studies revealed high-field properties of the related molybdates Ni$_2$Mo$_3$O$_8$ \cite{tang2021metamagnetic} and Co$_2$Mo$_3$O$_8$ \cite{szaller2022coexistence, tang2022successive}. 
Owing to the weaker spin anisotropy, applying high magnetic fields along the $c$ axis causes spin-flop and stabilizes canted spin states in these materials. 
Strong spin anisotropy in Fe$_2$Mo$_3$O$_8$ originated from strong spin-lattice coupling realizes unusual multistep magnetoelectric plateaux and 3-5 times larger change in the electric polarization by the transition from the AFM to FSP state.

In summary, we report unusual multistep magnetization and the corresponding giant magnetoelectric effect in single-crystalline Fe$_2$Mo$_3$O$_8$.
A fully spin-polarized state was realized in this system in magnetic fields of up to 66 T applied along the polar axis at 1.4 K. 
This multistep magnetization process is interpreted using a combined triangular lattice model with strong Ising-type anisotropy. 
The exchange striction mechanism reasonably reproduces the corresponding multistep electric polarization.

%

\newpage

\begin{figure}[!ht]
    \centering
    \includegraphics[scale=0.6]{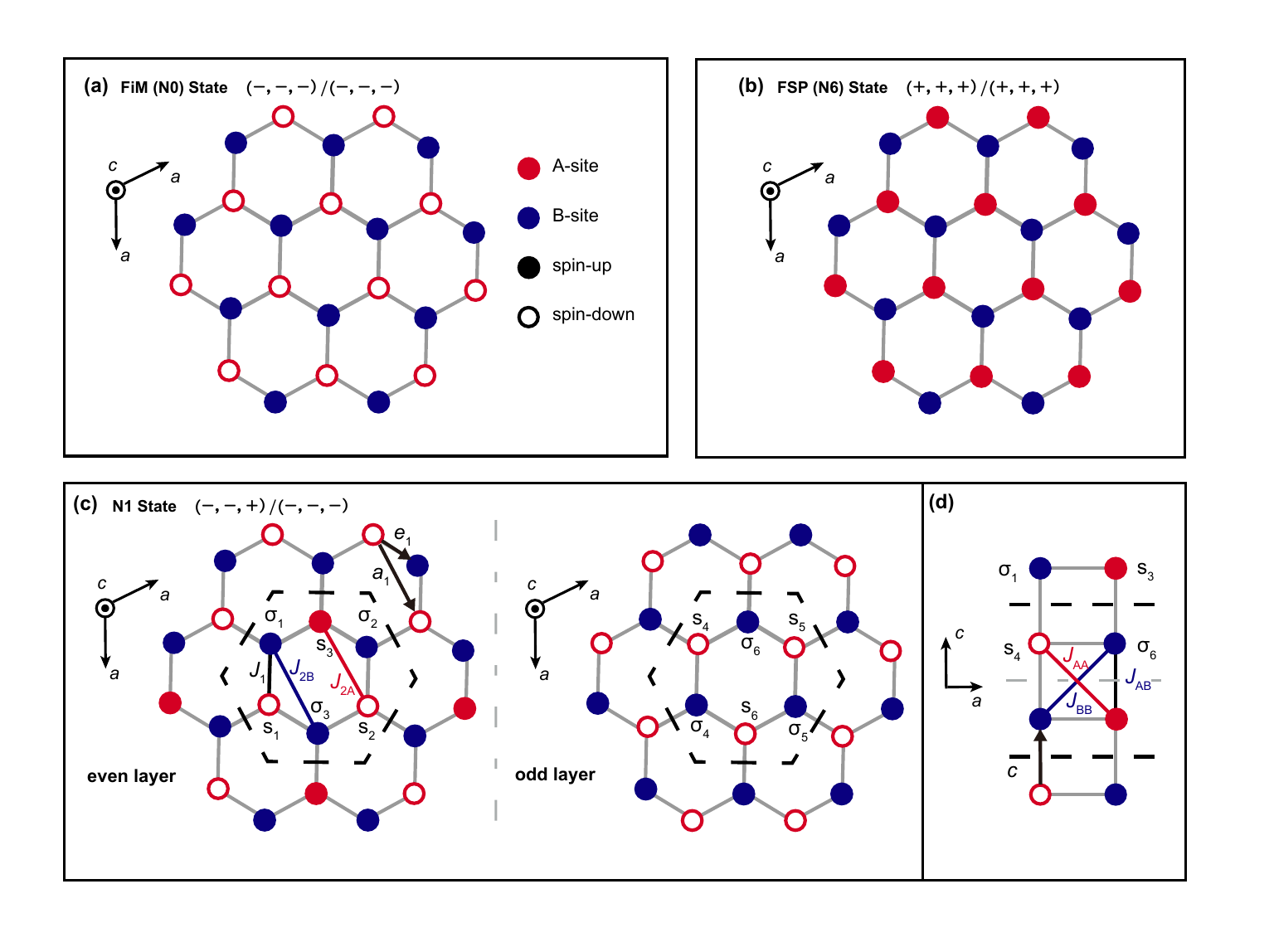} 
    \caption{ 
        (a,b) Schematic illustrations of the FiM and FSP states in the $ab$ plane, respectively. 
        (c,d) Schematic illustrations of the N1 state projected on the $ab$ and $ac$ planes, respectively.
        The thick broken lines represents the boundaries in magnetic unit cells. 
    }   
    \label{fig:fig4}
    \end{figure}

\begin{figure}[!ht]
    \centering
    \includegraphics[scale=1.2]{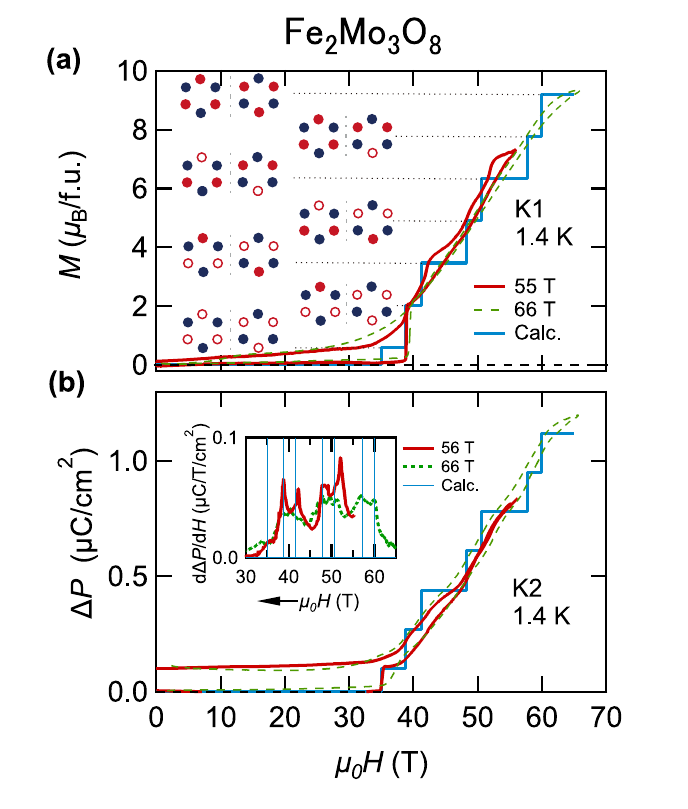} 
    \caption{
        Comparison between the experimental (red, green) and calculated (blue) curves of (a)$M$ and (b)$\Delta P$.
        Here, $\Delta P$ represents the change from the AFM state. 
        The inset in (b) shows the field derivative of experimental and calculated electric polarization curves. 
        The black arrow indicates that the field derivative curves in the inset are derived from the field-decreasing process.
    }   
    \label{fig:fig5}
    \end{figure}

\end{document}